\documentclass[prl,showpacs,superscriptaddress,floatfix,twocolumn]{revtex4}
\usepackage[latin1]{inputenc}
\usepackage[english]{babel}
\usepackage{graphicx}
\usepackage{amsmath}
\usepackage{amssymb}
\usepackage{amscd}
\usepackage{eucal}
\usepackage{color}
\usepackage{bm}
\usepackage{ulem}
\input {epsf}
\begin{document}

\title{{Gate-controlled Persistent Spin Helix State in Materials with \\ Strong Spin-Orbit Interaction}}
\author{M. Kohda}
\affiliation{Department of Materials Science, Tohoku University, Sendai 980-8579, Japan}
\affiliation{PRESTO, Japan Science and Technology Agency, Saitama 332-0012, Japan}
\author{V. Lechner}
\affiliation{Institut f\"ur Experimentelle und Angewandte Physik, Universit\"at Regensburg, D-93040 Regensburg, Germany}
\author{ Y. Kunihashi}
\affiliation{Department of Materials Science, Tohoku University, Sendai 980-8579, Japan}
\author{T. Dollinger}
\affiliation{Institut f\"ur Theoretische Physik, Universit\"at Regensburg, D-93040 Regensburg, Germany}
\author{P. Olbrich}
\affiliation{Institut f\"ur Experimentelle und Angewandte Physik, Universit\"at Regensburg, D-93040 Regensburg, Germany}
\author{C. Sch\"{o}nhuber}
\affiliation{Institut f\"ur Experimentelle und Angewandte Physik, Universit\"at Regensburg, D-93040 Regensburg, Germany}
\author{I.~Caspers}
\affiliation{Institut f\"ur Experimentelle und Angewandte Physik, Universit\"at Regensburg, D-93040 Regensburg, Germany}
\author{V.V.~Bel'kov}
\affiliation{ Ioffe Physical-Technical Institute, 194021 St.\,Petersburg, Russia}
\author{L.E. Golub}
\affiliation{ Ioffe Physical-Technical Institute, 194021 St.\,Petersburg, Russia}
\author{D. Weiss}
\affiliation{Institut f\"ur Experimentelle und Angewandte Physik, Universit\"at Regensburg, D-93040 Regensburg, Germany}
\author{K. Richter}
\affiliation{Institut f\"ur Theoretische Physik, Universit\"at Regensburg, D-93040 Regensburg, Germany}
\author{J. Nitta}
\affiliation{Department of Materials Science, Tohoku University, Sendai 980-8579, Japan}
\author{S.D. Ganichev}
\email{sergey.ganichev@physik.uni-r.de}
\affiliation{Institut f\"ur Experimentelle und Angewandte Physik, Universit\"at Regensburg, D-93040 Regensburg, Germany}
\date{\today }

\begin{abstract}
In layered semiconductors with  spin-orbit interaction (SOI) a persistent spin helix (PSH) state with suppressed spin relaxation is expected if the strengths of the Rashba and Dresselhaus SOI terms, $\alpha$ and $\beta$, are equal. 
Here we demonstrate gate control and detection of the PSH in two-dimensional electron systems with strong SOI including terms cubic in momentum. 
We consider strain-free InGaAs/InAlAs quantum wells and first determine a ratio $\alpha/\beta\!\simeq\! 1$ for non-gated structures by measuring the spin-galvanic and circular photogalvanic effects.  
Upon gate tuning the Rashba SOI strength in a complementary magneto-transport experiment, we then monitor the complete crossover from weak antilocalization via weak localization to weak antilocalization, where the emergence of weak localization reflects a PSH type state.
A corresponding numerical analysis reveals that such a PSH type state indeed prevails even in presence of strong cubic SOI, however no longer at $\alpha=\beta$.
\end{abstract}

\date{\today}


\maketitle

An electron moving in an electric field experiences, in its rest frame, an effective magnetic field pointing perpendicularly to its momentum.
The coupling of the electron's spin to this magnetic field is known as spin-orbit interaction.
The ability to control the corresponding magnetic field, and thereby spin states, all electrically in gated semiconductor heterostructures~\cite{Nitta97,Engels97}  is a major prerequisite and motivation for research towards future semiconductor spintronics.
However, on the downside, the momentum changes of an electron moving through a semiconductor cause sudden changes in the magnetic field leading to spin randomization. 
Hence, suppression of spin relaxation in {the} presence of strong, tunable SOI is a major challenge of semiconductor spintronics.

In III-V semiconductor heterostructures two different types of SOI exist:
i) Rashba SOI~\cite{Rashba60}, originating from structure inversion asymmetry (SIA), is linear in momentum $\bm k$ with a strength $\alpha$ that can be controlled by an electric gate;
ii) Dresselhaus SOI~\cite{Dresselhaus55} due to bulk inversion asymmetry (BIA), which gives rise to a band spin splitting, given by $\bm k$-linear and $\bm k$-cubic contributions~\cite{footnote0}.
The strength of the linear in $\bm k$ term $\beta \!=\! \gamma \langle  k_z^2 \rangle$ (where $\gamma$ is a material  parameter) can hardly be changed as it stems from crystal fields.
These various spin-orbit terms in layered semiconductors are  described by the Hamiltonian $ H_{\rm SO} \!=\! H_{\rm R} \! + \! H_{\rm D}$ with Rashba and Dresselhaus terms
\begin{eqnarray}
 H_{\rm R}  & = &   \alpha\left(k_y \sigma_x - k_x \sigma_y\right) \, ,
\label{eq:Rashba} \\
 H_{\rm D} & = &   \beta\left(k_x \sigma_x - k_y \sigma_y\right) +
              \gamma\left(-\sigma_x k_x k_y^2 + \sigma_y k_y k_x^2 \right)  
\label{eq:Dresselhaus}
\end{eqnarray}
with $\sigma_x,\sigma_y$ the Pauli spin matrices \cite{Fabian}.

If the $\bm k$-cubic terms can be neglected, a special situation emerges if Rashba and Dresselhaus SOI are of equal strength:
$\alpha = \pm\beta$. Then spin relaxation is suppressed~\cite{Averkiev1999p15582,Schliemann03}. A co-linear
alignment of Rashba and Dresselhaus effective magnetic fields gives rise to spin precession around a fixed axis, leading to spatially periodic modes~\cite{Schliemann03} referred to as persistent spin helix (PSH)~\cite{Bernevig06} and reflecting the underlying $SU(2)$ symmetry in this case. 
The PSH is robust against all forms of spin-independent scattering. 
This favorable situation where spin relaxation is suppressed while the spin degree of freedom is still susceptible to electric fields has led to various theoretical proposals for future spintronics settings~\cite{Schliemann03,Cartoixa03,Nitta2012} that are based on adjusting $\alpha=\beta$ by tuning $\alpha$ through an electric gate.

Experimentally, the existence of the PSH has been demonstrated by means of transient spin-gating spectroscopy~\cite{Koralek09} in ungated GaAs/AlGaAs quantum wells. 
While the weak $\bm k$-cubic SOI in this experiment barely affects the PSH formation, the important question arises whether a PSH type state will generally survive in materials with strong SOI where finite $\bm k$-cubic terms gain importance, in particular for heterostructures at higher charge carrier densities.
Also, compared to the linear case, much less is known theoretically~\cite{Glazov06,Stanescu07,Duckheim10,Luffe11} about the robustness of the PSH in this general case.

In this Letter we demonstrate in two independent, transport and optical, experiments the formation of a PSH state in a material with strong SOI, InGaAs quantum wells. 
On the one hand, we consider quantum corrections to the magneto-conductance to detect the PSH: 
While SOI generally leads to spin randomization and thereby to weak antilocalization (WAL)~\cite{Knap96}, systems with linear SOI obeying $\alpha = \pm \beta$, where spins are not rotated along closed back-scattered trajectories, should exhibit weak localization (WL)~\cite{Pikus95,Zaitsev05,Glazov06,Kettemann11}. 
By electrically tuning the Rashba SOI in InGaAs samples we monitor a crossover from WAL to WL and back, thereby identifying a PSH type  state even in the presence of $\bm k$-cubic SOI~\cite{footnote1}.
In order to get independent information on the ratio $\alpha/\beta$ we performed complementary measurements for the same QW {employing} the spin-galvanic (SGE)~\cite{Ganichev2,Belkov08} and circular photogalvanic (CPGE)~\cite{Ganichev3} effects that are insensitive to certain cubic spin-orbit terms. 
Comparison with the weak localization experiments hence enables us to extract information on the role of the cubic terms. In a corresponding numerical analysis
we show that a PSH type state indeed remains for finite cubic SOI.
However, this happens at $\alpha \neq \beta$ even if $\beta$ is renormalized by $\bm k$-cubic SOI. 

\begin{figure}
 \centering
  \includegraphics[width=1.0\columnwidth]{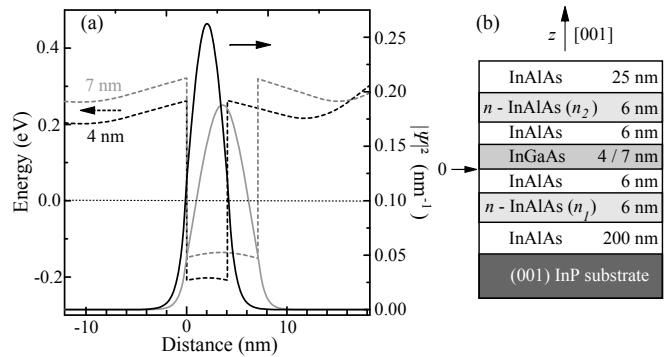}
\caption{
  (a) Energy-band profile of conduction band
{with the Fermi energy at $E_{\rm F} = 0$} in ${\rm In_{0.53}Ga_{0.47}As / In_{0.52}Al_{0.48}As}$ 2DEG structures calculated by using a Poisson-Schr\"{o}dinger equation solver. 
Black and gray lines correspond to 4 and 7 nm samples, respectively. 
Full lines denote the normalized electron probability density $|\psi|^2$.
(b) Schematic cross section of the structures under study.
}
\label{figure01}
\end{figure}

We experimentally investigate the PSH state in strain-free (001)-grown ${\rm In_{0.53}Ga_{0.47}As / In_{0.52}Al_{0.48}As}$ quantum well (QW) structures hosting a two-dimensional electron gas~(2DEG).
The QWs were designed to achieve almost equal linear Rashba and Dresselhaus coefficients, $\alpha$ and $\beta$, at zero gate voltage. Since $\beta$ is usually much smaller than $\alpha$ in InGaAs 2DEGs~\cite{Faniel11}, we needed to enhance $\beta$ and to reduce the built-in Rashba SOI. 
Since in QW structures $\beta \propto \langle k^2_z \rangle \propto 1/L_W^2$, we designed sufficiently narrow QWs of width $L_W = 4${~nm} and 7~nm. 
Furthermore, $\alpha$ was designed to be small at zero gate bias by preparing symmetric InGaAs QWs by placing two {Si doping} layers with densities $n_1 = 1.2$  and $n_2 = 3.2 \times 10^{18}~{\rm cm}^{-3}$ into the InAlAs barriers, {each placed 6~nm away} from the QW [see Fig.~\ref{figure01}(b)]. 
Here, the higher doping level on the top side of the QW compensates the surface charges. 
Figure~\ref{figure01}(a) shows the {resulting} conduction band structure and the {electron distribution} of {the} 4~nm and 7~nm ${\rm In_{0.53}Ga_{0.47}As / In_{0.52}Al_{0.48}As}$ QW structures.
For the gate control of the PSH state, the epitaxial wafers were processed into $20~\mu$m~$\times$~$80~\mu$m Hall bar structures with an ${\rm Al_2 O_3}$ gate insulator and a Cr/Au top gate.

\begin{figure}
 \centering
  \includegraphics[width=1.0\columnwidth]{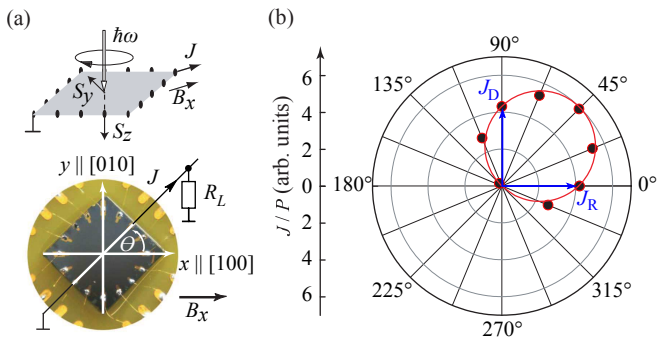}
\caption{
(Color Online)
(a) Sketch of the experimental arrangement (top) and sample geometry used for measurements of the spin-galvanic effect. 
Here we used circularly polarized light at normal incidence {with the} magnetic field applied along {the} $x \parallel [100]$-direction.
The photocurrent $J(\theta)$ is mapped by measuring {successively} signals from opposite contact {pairs}. 
(b)  {Azimuthal} dependence of the SGE current $J$ measured in {a} 4~nm QW at room temperature, $\lambda = 148$~$\mu$m and at $B_x = 0.8$~T. 
The solid line shows the fit according to $J = J_{\rm R} \cos\theta + J_{\rm D} \sin\theta $ with the ratio of $ J_{\rm R} / J_{\rm D} = 0.98~\pm~0.08$.
}
\label{figure02}
\end{figure}

The photocurrent experiments were carried out with terahertz radiation of a  pulsed ${\rm NH_3}$ laser~\cite{JETP1982} with peak power of 5 to 10~kW, controlled by a reference photon drag detector~\cite{JTPL1985}.
Our laser generates single 100~ns pulses with a repetition rate of 1~Hz and wavelengths $\lambda =$ 90.5, 148 or 280~$\mu$m. 
The corresponding photon energies are 13.7, 8.4 and 4.4~meV, respectively, much smaller than energy gap and subband separation. 
Therefore, the absorption is due to indirect transitions only within the lowest conduction subband.  
The configuration used for the spin-galvanic effect experiments is sketched in Fig.~\ref{figure02}(a).
For these measurements the samples  were irradiated by circularly polarized light along the growth direction ($z$-axis), and an external magnetic field with strengths up to 1~T was applied along the [100]-axis. 
The light generates a nonequilibrium spin polarization $\bm S \parallel z$ which, by means of the in-plane magnetic field, can be rotated into the QW plane.
A spin-galvanic current $\bm J$, flowing due to asymmetric spin relaxation~\cite{Ganichev2}, was measured [see  Fig.~\ref{figure02}(a)] via {the }voltage drop across a 50~$\Omega$ load resistor.
For these experiments, carried out at temperatures $T= 5$~K and 296~K, the ungated samples,  made of the same batches as the Hall bar structures, were used.

Figure~\ref{figure02}(b) shows the signal of the spin-galvanic effect measured at room-temperature for the 4~nm QW along different in-plane directions, determined by the angle $\theta$ with respect to the fixed in-plane magnetic field ${\bm B} \parallel x$.
The current component, $J_R$, parallel to the magnetic field is driven by the Rasba spin splitting, while the perpendicular component, $J_D$ is caused by the Dresselhaus SOI~\cite{Ganichev2,Ganichev3}. 
The data presented in Fig.~\ref{figure02}(b) can be well fitted by $J = J_{\rm R} \cos\theta + J_{\rm D} \sin\theta $, with $ J_{\rm R} / J_{\rm D} = 0.98 \pm 0.08$. 
This ratio is related to that between the linear Rashba and Dresselhaus SOI strengths, $ J_{\rm R} / J_{\rm D} = \alpha / \tilde{\beta}$.
The renormalized coefficient $\tilde{\beta}$ is described by Eq.~(\ref{eq:omega-1}) below.

The room-temperature results indicate that our ungated 4~nm QW samples are in a regime where the formation of a PSH is expected.
At 5~K we find for the same sample $\alpha / \tilde{\beta}= {1.08~\pm~0.08}$ showing a weak temperature dependence of this ratio. 
For the 7~nm QW with smaller $\beta$ we find {similarly} $\alpha/\tilde{\beta} = 3.97~\pm~0.08$ and $4.00~\pm~0.08$ at $T =$~296 and 5~K, respectively.
Alternatively, the ratio between SOI strengths can {also} be extracted analyzing photocurrents arising {from} the circular photogalvanic effects~\cite{Ganichev3}. 
We obtain a ratio ${\alpha/\tilde{\beta}=1.04~\pm~0.07}$ for the ungated 4 nm sample at room temperature, in line with our SGE analysis.

The renormalized coefficient $\tilde{\beta}$ arises when decomposing $H_{\rm D}$ in Eq.~(\ref{eq:Dresselhaus}) [with  $\bm{\sigma} \!=\! (\sigma_x,\sigma_y)$]
into~\cite{Iordanskii94} 
\begin{equation}
 H_{\rm D} =  \hbar \, \bm{\sigma} \left(\bm{\Omega}_1^{\rm D} +  \bm{\Omega}_3^{\rm D}\right) \;
\label{eq:Dresselhaus2}
\end{equation}
with 
\begin{eqnarray}
    \hbar \bm{\Omega}_1^{\rm D} & = &  \tilde{\beta} \left(k_x \hat{\bm{x}} -k_y
\hat{\bm{y}}\right) \; ; \quad
\tilde{\beta} = \beta - \frac{\gamma}{4}\langle k^2\rangle \, ,
\label{eq:omega-1}
\\
\hbar \bm{\Omega}_3^{\rm D} & = &
- \frac{\gamma}{4} k^3 \left( \hat{\bm{x}} \cos3\vartheta_{\bm k} + \hat{\bm{y}}
\sin3\vartheta_{\bm k}  \right) \,  .
\label{eq:omega-3}
\end{eqnarray}
Here $k_x \!=\!k \cos \vartheta_{\bm k}, k_y \!=\! k \sin \vartheta_{\bm k}$ and $ \bm {\hat  x},~ \bm{\hat y}$ denote unit vectors.
The measured photo\-currents are related to the first-order harmonics ($\propto \sin{\vartheta_{\bm k}}$ and $\cos{\vartheta_{\bm k}}$ ) in the Fourier expansion of the nonequilibrium electron distribution function only~\cite{Ganichev2,Ganichev3}. 
Consequently, the photocurrent is proportional to the linear Rashba term $H_{\rm R}$, see Eq.~(\ref{eq:Rashba}), and renormalized Dresselhaus term $\hbar \bm{\sigma}\bm{\Omega}_1^{\rm D}$, see Eq.~(\ref{eq:omega-1}), but it is insensitive to the third harmonic of the cubic SOI term given by Eq.~(\ref{eq:omega-3}).

In a second, complementary, transport experiment we measured the quantum correction to the magneto-conductivity in the gated Hall bar structures in the presence of an external magnetic field $\bm B$, pointing perpendicularly to the QW plane~\cite{Sch08}.
At $T =$  1.4~K, various magneto-conductivity profiles were recorded for different strengths of the Rashba SOI by varying the gate voltage $V_{\rm g}$. 

The carrier density $N_s(V_{\rm g})$, the mobility $\mu (V_{\rm g})$ and the mean free path $l (V_{\rm g})$ were extracted from sheet resistivity and  periodicity of the Shubnikov-de Haas oscillations giving for the 4 nm (7~nm) sample $N_s = 4.12 \times 10^{12}$ ($3.48\times 10^{12}$)~cm$^{-2}$, $\mu = 15000$ (27000)~cm$^2$/Vs  and $l = 0.50$ (0.83)~$\mu$m at $V_g = 0$~V and $T = 1.7$~K.

\begin{figure}
 \centering
  \includegraphics[width=1.00\columnwidth]{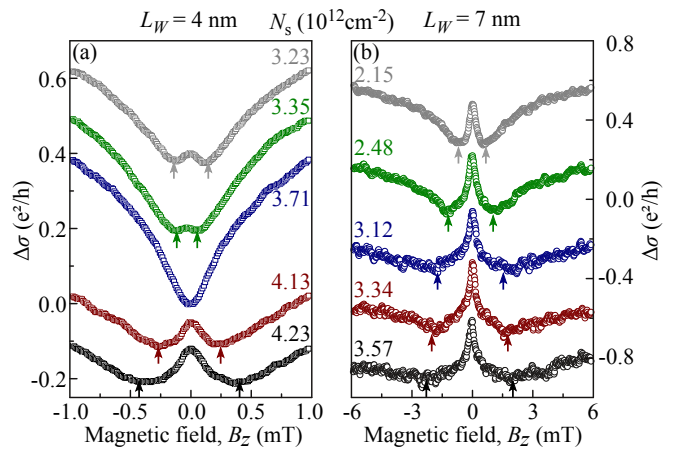}
\caption{
(Color Online) Magneto-conductance profiles (in units of $e^2/h$) measured at different gate voltages, i.e., carrier densities $N_s$, for (a) 4~and  (b) 7~nm QWs at $T =$ 1.4~K. 
All curves in (a) and (b) are shifted positively (gray and green) and negatively (black and red) with respect to the blue curve, for which $\Delta \sigma = 0$ at $B_z = 0$~mT.
For the 4~nm QW a clear WL dip occurs for a carrier density of $3.71 \times 10^{11}$~cm$^{-2}$, which is absent for the 7~nm QW.
}
\label{figure03}
\end{figure}

Figures~\ref{figure03}(a) and (b) show the measured magneto-conductance profiles at different gate voltages for the 4~nm and 7~nm wide QWs, respectively.
On the one hand, for the 7~nm QW, only  WAL characteristics are observed, which get enhanced with increasing $N_s$.
On the other hand, most notably, the magneto-conductance for the 4~nm QW near $B\!=\!0$ changes from WAL to WL characteristics and back again to WAL upon increasing $N_s$ from 3.23 to 4.23$\times 10^{12}$~cm$^{-2}$. 
The occurrence  of WL (at  $N_s \! = 3.71\times 10^{12}$~cm$^{-2}$) reflects suppressed spin relaxation, and the observed sequence WAL-WL-WAL unambiguously
indicates that -- even in presence of strong $\bm k$-cubic SOI -- a PSH condition is fulfilled in the WL region.

Since the band profiles of the 4~nm and 7~nm wide quantum wells are very similar (see Fig.~\ref{figure01}) we expect comparable values of the Rashba SOI. 
The difference in {the bias dependence of} the magneto-conductance of the two devices we hence ascribe to the different Dresselhaus SOI strengths $\beta \propto 1/L_W^2$.
While for the 7~nm sample at zero bias the difference between $\alpha$ and a small $\beta$ is too large to get tuned to comparable values, the larger $\beta$ of the 4~nm QW {enables} gate voltage tuning of $\alpha$ to meet the condition for PSH formation, $\alpha \simeq \beta$, associated with the occurrence of the  WL peak.

%

\begin{figure}
 \centering
\includegraphics[width=0.90\columnwidth]{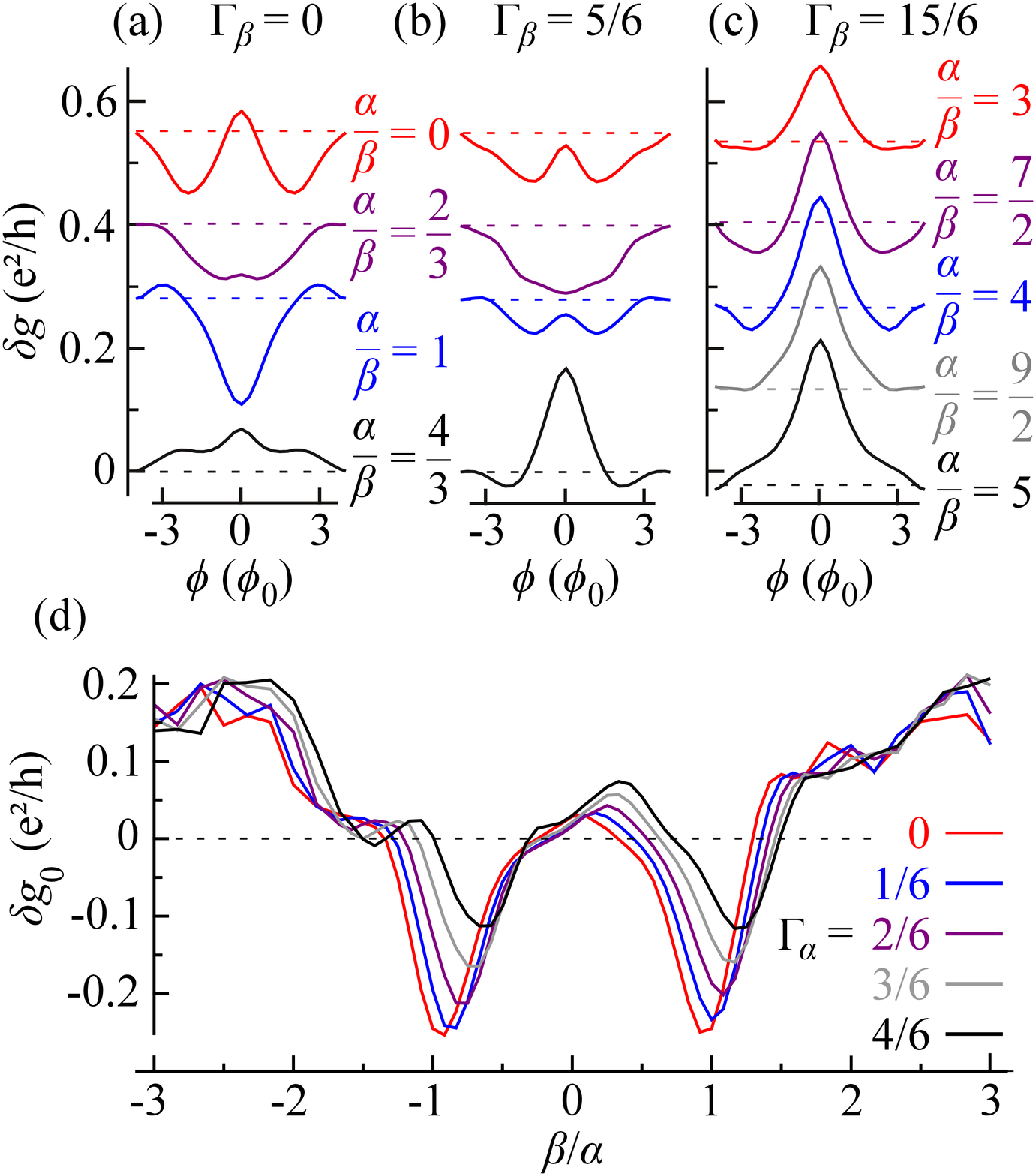}
\caption{
Numerical analysis of the PSH: Panels (a,b): Profiles of the average conductance correction $\delta g(\phi) = g(\phi)-g(\phi_c)$ (in units of $e^2/h$) as function of flux $\phi$ (normalized by the flux quantum $\phi_0$) for disordered mesoscopic conductors with SOI ratios  $\alpha/\beta=$ 0 to 4/3, see text.
(a): Case of vanishing cubic SOI, $\Gamma_{\beta}= 0$:
weak antilocalization is observed due to spin interference except for $\alpha/\beta=1$ when spin relaxation is suppressed by exact $SU(2)$ symmetry.
(b): A finite cubic term $\Gamma_{\beta}=5/6$ leads to a weak localization signature at a shifted value $\alpha/\beta<1$, in agreement with the experimental observations in Fig.~\ref{figure03}(a).
(c): Absence of weak localization for increased  $\Gamma_{\beta}=16/5$ and larger ratios $\alpha/\beta$, consistent with the increase of the corresponding quantities when changing $L_W\!=\!$~4~nm [Fig.~\ref{figure03}(a)] to 7~nm [Fig.~\ref{figure03}(b)]. 
Up to the lowest curves all data have been shifted by arbitray offsets for the sake of clarity.
(d): Magneto-conductance correction $\delta g_0= \delta g(\phi\!=\!0)$ as a function of the ratio $\beta/\alpha$ for fixed $\alpha$ and various values of $\Gamma_{\alpha}=k_{\rm F}^2 \gamma/\alpha$, WL dips are linearly shifted with increasing $\Gamma_{\alpha}$ indicating a modified PSH condition in $\alpha/\beta$
space.
}
\label{figure04}
\end{figure}
%

While the photocurrent experiments are insensitive to the third harmonic of the cubic term, Eq.~(\ref{eq:omega-3}),  and thus reveal the ratio $\alpha / \tilde{\beta}$, the transport experiment probes spin randomization due to the entire SOI contribution, Eqs.~(\ref{eq:Rashba}) and~(\ref{eq:Dresselhaus}). 
Hence the key question remains, namely: 
What is the general condition for the appearance of a PSH state, respectively WL, replacing the  one, $\alpha=\beta$, for the linear case? 
To answer this question, and for an in-depth analysis of the transport experiments, we systematically studied the numerically computed crossover from WAL to WL in the magneto-conductance of disordered conductors under variation of the quantities $\alpha, \beta, \gamma$ treated as independent parameters in our model (thus neglecting the connection of $\beta$ and $\gamma$ in realistic systems with fixed $L_W$). 
Our tight-binding calculations are based on an efficient recursive Green function algorithm~\cite{Wimmer08} within the Landauer formalism. 
We consider a diffusive, two-dimensional phase-coherent mesoscopic conductor with periodic boundary conditions perpendicular to the transport direction. 
We employ the full SOI Hamiltonian, Eqs.~(\ref{eq:Rashba}) and~(\ref{eq:Dresselhaus}), plus the contribution from an external perpendicular $\bm B$-field. 
While we cannot model the parameters of the 2D bulk experiment, since we are numerically limited to energy scales smaller than the realistic Fermi energies, we chose system sizes as well as energies and SOI strengths such that the ratio of the relevant parameters are comparable to experiment.
These parameters consist of the ratios  $\alpha/\beta$ and $\Gamma_{\alpha(\beta)}= \gamma k_{\rm F}^2/\alpha(\beta)$ which quantify the relative strength in {the} Hamiltonian (\ref{eq:Rashba},\ref{eq:Dresselhaus}) of cubic and linear SOI at a given Fermi momentum $k_{\rm F}$.
By averaging over typically 100 disorder realizations we then computed the quantum (WL and WAL) correction
$\delta g(\phi) \!=\! \langle g(\phi) \rangle  - \langle g(\phi_c)\rangle\!\!$
\, by subtracting the average conductance $\langle g(\phi_c) \rangle$ at magnetic fields $B_c$ where coherent backscattering is suppressed from the average magneto-conductance $\langle g(\phi)\rangle$, with $\phi$ being the magnetic flux through the system generated by the field $\bm B$.

Our main numerical results are summarized in Fig.~\ref{figure04}.
Panels (a,b) show magneto-conductance profiles for various ratios of $\alpha/\beta$ for (a) vanishing and (b) finite $\Gamma_{\beta}$.
For the linear case, panel (a), we find, as expected, a WAL-WL-WAL crossover with pronounced WL dip for $\alpha/\beta\!=\!1$ reflecting the PSH.
For finite cubic SOI strength,  $\Gamma_{\beta}\!=\!5/6$, our numerics support the experimental findings [see Fig.~\ref{figure03}(a)] that the crossover prevails, although with less pronounced WL dip.
A rough estimate for the experiments with $L_W = 4$~nm leads to $\Gamma_{\beta}\approx 0.4$, which is close to the numerical $\Gamma_{\beta}$. 
 Figure~\ref{figure04}(c) is the numerical counterpart of Fig.~\ref{figure03}(b), and predicts the absence of weak localization in a regime of increased  $\Gamma_{\beta}$ and $\alpha/\beta$, where the factor by which $\Gamma_{\beta}$ is increased matches the value when the experiment at $L_W\!=\!$~4~nm [Fig.~\ref{figure03}(a)] is compared to 7~nm [Fig.~\ref{figure03}(b)].
Moreover, the strongest WL signal appears no longer at $\alpha/\beta\!=\!1$.
In order to explore a refined condition for PSH behavior, we present in Fig.~\ref{figure04}(d) $\delta g(\phi=0)$ as a function of $\beta/\alpha$ for fixed $\alpha$ and stepwise increasing values of $\Gamma_{\alpha}$ from $0$ to $4/6$. 
All curves display regimes of both WL and WAL behavior. We find two trends for increasing cubic SOI:
(i) The WL dips diminish indicating the onset of spin relaxation, i.e., breaking of the exact PSH $SU(2)$ symmetry of the linear case.
(ii) The conductance minima (WL) arising at $\beta/\alpha\!=\!\pm 1$ for $\Gamma_{\alpha} \!=\! 0$ are linearly shifted towards larger values of $|\beta/\alpha|$ with growing $\Gamma_{\alpha}$.
 We find  a shift $\sim 0.4  \gamma \langle k^2 \rangle$ that differs from the shift $(1/4) \gamma \langle k^2 \rangle$ entering into $\tilde{\beta}$ in Eq.~(\ref{eq:omega-1}). 
Hence our numerical analysis indicates that the condition for PSH formation deviates from common assumption $|\alpha|\! =\! |\tilde{\beta}|$. While often neglected, the cubic Dresselhaus terms (\ref{eq:omega-3}) not only speed up spin relaxation but furthermore move the PSH point in three-dimensional SOI parameter space.  
This can also explain the small difference in the carrier density for observed WL-like behavior in transport ($N_s  = 3.71~\times~ 10^{12}~{\rm cm}^{-2}$)
and for $|\alpha| = |\tilde{\beta}|$ for the photocurrents ($N_s \sim  4.12~ \times~ 10^{12}~{\rm cm}^{-2} $) and suggests a refined condition for PSH-type behavior in the presence of $\bm k$-cubic terms.

In conclusion, we have demonstrated   the existence and gate control of a PSH state in an InGaAs/InAlAs QW structure with strong, $\bm k$-cubic SOI by precise  engineering the Rashba and Dresselhaus SOIs.
The corresponding spin splittings have been deduced by utilizing methods based on the study of the anisotropies in WAL, SGE and CPGE analysis.
By applying a gate electric field, a clearcut WAL-WL-WAL transition was observed. 
The results obtained by the different independent experimental techniques are in a good agreement, also with quantum transport calculations, and both experiments and theory reveal the robustness of the PSH.
We thereby demonstrate that this state can be achieved even in structures with strong SOI and a substantial $\bm k$-cubic SOI contribution.
The essential prerequisite is that for zero bias voltage $\alpha$ and $\beta$ are close to each other, a condition which can be reached in very narrow and almost symmetric QWs due to a specially designed doping profile.
However, in contrast to systems with dominating $\bm k$-linear spin splitting, the PSH is obtained for close, but nonequal Rashba and Dresselhaus strengths.

We thank M.M. Glazov for fruitful discussions.
This work was partly supported by Grants-in-Aids from JSPS, MEXT, Japanese-German joint DFG research unit FOR 1483, Elitenetzwerk Bayern. 
LG thanks RFBR, Russian President grant for young scientists, and EU program ``POLAPHEN'' for support.

\end{document}